\documentclass[preprint,aps,prl,amssymb]{revtex4}
\usepackage{graphicx}
\usepackage{amsmath}
\usepackage{amssymb}
\usepackage{color}

\begin{document}
\title{Pressure and isotope effect on the anisotropy of MgB$_{2}$}

\author{T. Schneider$^{\text{1}}$ and D. Di Castro$^{\text{1, 2}}$}
\affiliation{$^{\text{(1)}}$ Physik-Institut der Universit\"{a}t Z\"{u}rich,\\
Winterthurerstrasse 190,\\ CH-8057 Z\"{u}rich, Switzerland\\
$^{\text{(2)}}$ INFM-Coherentia and Dipartimento di Fisica,
Universita' di Roma "La Sapienza",\\ P.le A. Moro 2,\\ I-00185
Roma, Italy}
\date{\today}

\begin{abstract}
We analyze the data for the pressure and boron isotope effect on
the temperature dependence of the magnetization near $T_{c}$.
Invoking the universal scaling relation for the magnetization at
fixed magnetic field it is shown that the relative shift of
$T_{c}$, induced by pressure or boron isotope exchange, mirrors
essentially that of the anisotropy. This uncovers a novel generic
property of anisotropic type II superconductors, inexistent in the
isotropic case. For MgB$_{2}$ it implies that the renormalization
of the Fermi surface topology due to pressure or isotope exchange
is dominated by a mechanism controlling the anisotropy.
\end{abstract}
\maketitle

It is well documented that both, hydrostatic pressure\cite{tomita}
and boron isotope exchange\cite{hinks,daniele} lead in MgB$_{2}$
to a reduction of the transition temperature $T_{c}$. To be
specific, the dependence of $T_{c}$ on hydrostatic (He-gas)
pressure has been determined to $1$ GPa. $T_{c}$ was found to
decrease linearly and reversibly under pressure at the rate
$dTc/dP=-1.11\pm $ $0.02$ K/GPa\cite{tomita}, consistent with
$dTc/dP=-1.24\pm $ $0.05$ K/GPa reported by Di Castro \textit{et
al}.\cite {dicastro}. This corresponds with $T_{c}=T_{c}\left(
P=0\right) =39.3$ K to the relative change $\Delta
T_{c}/T_{c}\simeq -0.04$ at $P=1.13$GPa. Nearly the same value,
$\Delta T_{c}/T_{c}\approx -0.03$ was derived from the change of
the magnetization upon boron isotope exchange\cite{hinks,daniele}.
Although the anisotropy of MgB$_{2}$ is moderate, the
compressibility along the $c$-axis is significantly (64\%) larger
than that along the $a$-axis\cite {tomita}. Thus, the binding
within the boron layers is stronger than between the layers.
Reversible torque\cite{angst} and magnetization measurements\cite
{kim} near $T_{c}$ also revealed anisotropic superconducting
properties characterized by $\gamma \left( T_{c}\right) =\xi
_{a}/\xi _{c}=\lambda _{c}/\lambda _{a}\approx 2$, where $\xi
_{a,c}$ denote the correlation lengths and $\lambda _{a,c}$ the
magnetic penetration depths along the $a$ - and $c$ - axis. Since
in a superconductor increasing anisotropy drives a 3D to 2D
crossover, enhances thermal fluctuations and reduces $T_{c}$, the
anisotropy is an important characteristic both for the basic
understanding of superconductors and for applications. For
example, Dahm and Schopohl\cite {dahm} calculated $\gamma $ in the
clean limit based on a detailed modelling of the electronic
structure that took into account the Fermi surface topology and
the two-gap nature of the mean-field order parameter. It was shown
that the strong temperature dependence of the anisotropy can be
understood as an interplay of the dominating gap in the $\sigma$
band, which possesses a small $c$-axis component of the Fermi
velocity, with the induced superconductivity on the $\pi $-band
possessing a large $c$-axis component of the Fermi velocity.
Furthermore, the anisotropy strongly affects the pinning and
critical currents.

Here, we concentrate on the behavior near $T_{c}$, where thermal
fluctuations must be taken into account, as evidenced by the
excess specific heat\cite{park} and the
magnetoconductivity\cite{kang}. Noting that these fluctuations
mediate universal behavior, e.g. a universal relation between
magnetization, anisotropy $\gamma $ and transition temperature
$T_{c}$, an analysis of the pressure and isotope effect on the
magnetization should uncover the universal behavior of anisotropic
superconductors and provide stringent constraints for microscopic
treatments. We analyze the data for the pressure\cite{dicastro}
and boron isotope\cite{hinks} effect on the temperature dependence
of the magnetization near $T_{c}$. Here an effective single gap
description is appropriate\cite{zhitomirsky}. When
three-dimensional (3D) Gaussian or 3D-XY thermal fluctuations
dominate, the combination $m\left( T,\delta ,H\right) /\left(
\gamma \epsilon ^{3/2}\left( \delta \right) T\sqrt{H}\right) $
adopts then at $T_{c}$ a fixed value \cite
{fisher,tshk,hubbard,x,babic,hoferdis,hofer,book,parks,tspstat,tsisom}
\begin{equation}
\frac{m\left( T_{c}\right) }{T_{c}\left[ \gamma \epsilon
^{3/2}\left( \gamma ,\delta \right) \right]
_{T_{c}}\sqrt{H}}=-\frac{k_{B}C}{\Phi _{0}^{3/2}}, \label{eq1}
\end{equation}
where $m=M/V$ is the magnetization per unit volume, $C$ a constant
adopting for Gaussian and 3D-XY fluctuations distinct universal
values. Furthermore, $\epsilon \left( \delta \right) =\left( \cos
^{2}\left( \delta \right) +\sin ^{2}\left( \delta \right) /\gamma
^{2}\right) ^{1/2}$, where $\delta $ is the angle between the
applied magnetic field $\mathbf{H}$ and the $c$-axis, $\Phi _{0}$
the flux quantum, and $k_{B}$ the Boltzmann constant. Thus,
plotting $m\left( T\right) /\left( \gamma \epsilon ^{3/2}\left(
\delta \right) \sqrt{H}\right) $ \textit{vs. }$T$, the data taken
in different fields cross at $T_{c}$. In powder samples this
relation reduces to
\begin{equation}
\frac{m\left( T_{c}\right) }{T_{c}\sqrt{H}f\left( \gamma \left(
T_{c}\right) \right) }=-\frac{k_{B}C}{\Phi _{0}^{3/2}},\text{
}f\left( \gamma \left( T_{c}\right) \right) =\left[ \gamma
\left\langle \epsilon ^{3/2}\left( \gamma ,\delta \right)
\right\rangle \right] _{T_{c}}.  \label{eq2}
\end{equation}

As the isotope and pressure effect on the magnetization at fixed
magnetic field is concerned it implies that the relative shifts of
magnetization per unit volume $m$, anisotropy $\gamma $ and
$T_{c}$ are not independent but related by .
\begin{equation}
\Delta m\left( T_{c}\right) /m\left( T_{c}\right) =\Delta M\left(
T_{c}\right) /M\left( T_{c}\right) -\Delta V\left( T_{c}\right)
/V\left( T_{c}\right) =\Delta f\left( \gamma \left( T_{c}\right)
\right) /f\left( \gamma \left( T_{c}\right) \right) +\Delta
T_{c}/T_{c}.  \label{eq3}
\end{equation}
On that condition it is impossible to extract these changes from
the temperature dependence of the magnetization. However,
supposing that close to criticality the magnetization data scale
within experimental error as
\begin{equation}
^{i}M\left( T\right) =\text{ }^{j}M\left( aT\right) ,  \label{eq4}
\end{equation}
where $^{i\neq j}M$ denotes the magnetization for different
isotopes or taken at different pressures, the universal relation
(\ref{eq3}) reduces to
\begin{equation}
-\frac{\Delta T_{c}}{T_{c}}=\Delta V\left( T_{c}\right) /V\left(
T_{c}\right) +\frac{\Delta f\left( \gamma \left( T_{c}\right)
\right) }{f\left( \gamma \left( T_{c}\right) \right) }=1-a.
\label{eq5}
\end{equation}
Hence, when Eq.(\ref{eq4}) holds true, the pressure and isotope
effect on $T_{c}$ mirrors that of the anisotropy $\gamma =\xi
_{ab}/\xi _{c}=$ $\lambda _{ab}/\lambda _{c}$ and of the volume.
In particular, when $\gamma \left( T_{c}\right) >>1$, $f\left(
\gamma \left( T_{c}\right) \right) \rightarrow 0.556\gamma \left(
T_{c}\right) $ and with that $-\Delta T_{c}/T_{c}=\Delta V\left(
T_{c}\right) /V\left( T_{c}\right) +\Delta \gamma /\gamma $.

\begin{figure}[tbp]
\centering
\includegraphics[height=5cm]{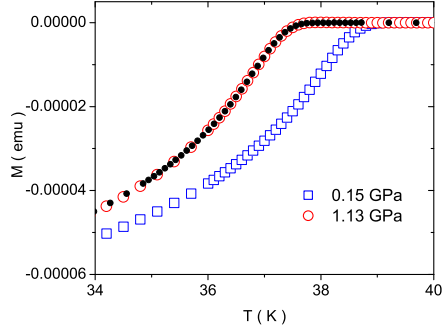}
\caption{Field cooled ($0.5$mT) magnetization of a MgB$_{2}$
powder sample \textit{vs}. $T$ near $T_{c}$ for $P=0.15$ and
$P=1.13$ GPa taken from Di Castro \emph{et
al}.\protect\cite{dicastro}. The dots are the $P=0.15$ data
rescaled according to Eq.(\ref{eq4}) with $a=0.968$.} \label{fig1}
\end{figure}

\begin{figure}[tbp]
\centering
\includegraphics[height=6cm]{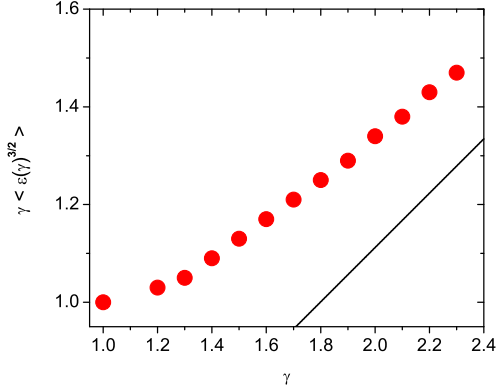}
\caption{$f\left( \gamma \right) =\gamma \left\langle \varepsilon
\left( \gamma \right) ^{3/2}\right\rangle $ \textit{vs}. $\gamma
$. The straight line indicates the asymptotic behavior in the
limit $\gamma \rightarrow \infty $, where $f\left( \gamma \right)
=\gamma \left\langle \left| \cos (\delta )\right|
^{3}\right\rangle \simeq 0.556\gamma $.} \label{fig2}
\end{figure}

We are now prepared to analyze the experimental data for the
pressure and isotope effect on the magnetization of MgB$_{2}$.
While the volume change upon isotope exchange is negligible, it
can be appreciable by applying pressure. In terms of the bulk
modulus $B\simeq 147.2$ GPa\cite{jorgensen} it is given by $\Delta
V/V=-\Delta P/B$ so that for $\Delta P\simeq 1$ GPa, $\Delta
V/V\simeq -0.007$. In Fig.\ref{fig1} we displayed the field cooled
($0.5$mT) magnetization of a MgB$_{2}$ powder sample \textit{vs}.
$T$ near $T_{c}$ for $P=0.15$ and $P=1.13$ GPa taken from Di
Castro \emph{et al}.\cite {dicastro}. The dots are the $P=0.15$
data rescaled according to Eq.(\ref {eq4}) with $a=0.968$. Noting
that the rescaled $P=0.15$ data collapse near $T_{c}$ within
experimental error onto the $P=1.13$ GP data, $\Delta M/M\simeq 0$
follows and Eq.(\ref{eq5}) applies as
\begin{equation}
-\frac{\Delta T_{c}}{T_{c}}=\Delta V\left( T_{c}\right) /V\left(
T_{c}\right) +\frac{\Delta f\left( \gamma \left( T_{c}\right)
\right) }{f\left( \gamma \left( T_{c}\right) \right) }=1-a\simeq
0.032. \label{eq6}
\end{equation}
To check the scaling analysis we note that the pressure dependence
of $\Delta T_{c}/T_{c}=\left( T_{c}\left( P\right) -T_{c}\left(
0\right) \right) /T_{c}\left( 0\right) $ is well described by
$\Delta T_{c}/T_{c}=-0.032P$ with $P$ in GPa\cite{dicastro},
yielding for $\Delta P=0.98$ GPa the value $\Delta
T_{c}/T_{c}=-0.032$, in excellent agreement with our estimate. In
addition to it, the scaling analysis reveals that the pressure
induced reduction of $T_{c}$ is mainly due to the anisotropy.
Indeed, since $\Delta T_{c}/T_{c}\simeq 4.5\Delta V\left(
T_{c}\right) /V\left( T_{c}\right) \simeq $ $-0.032$, there is a
significantly larger and positive anisotropy contribution $\Delta
f\left( \gamma \left( T_{c}\right) \right) /f\left( \gamma \left(
T_{c}\right) \right) \simeq 0.04$. Noting that $f\left( \gamma
\right) $, displayed in Fig.\ref{fig2}, increases with $\gamma $
the anisotropy is found to increase with pressure.

\begin{figure}[tbp]
\centering
\includegraphics[height=6cm]{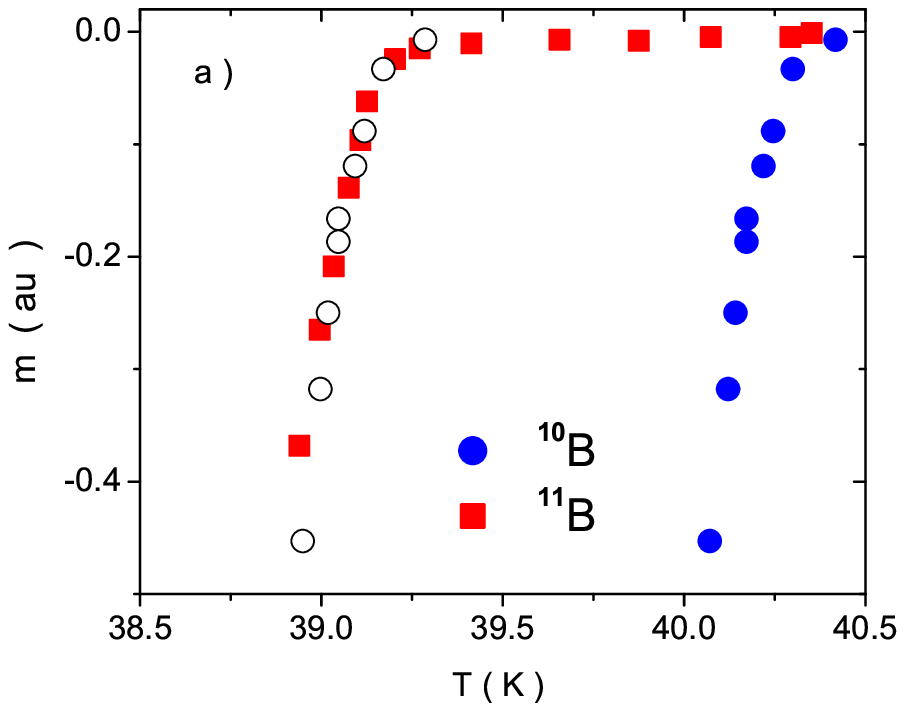}
\includegraphics[height=6cm]{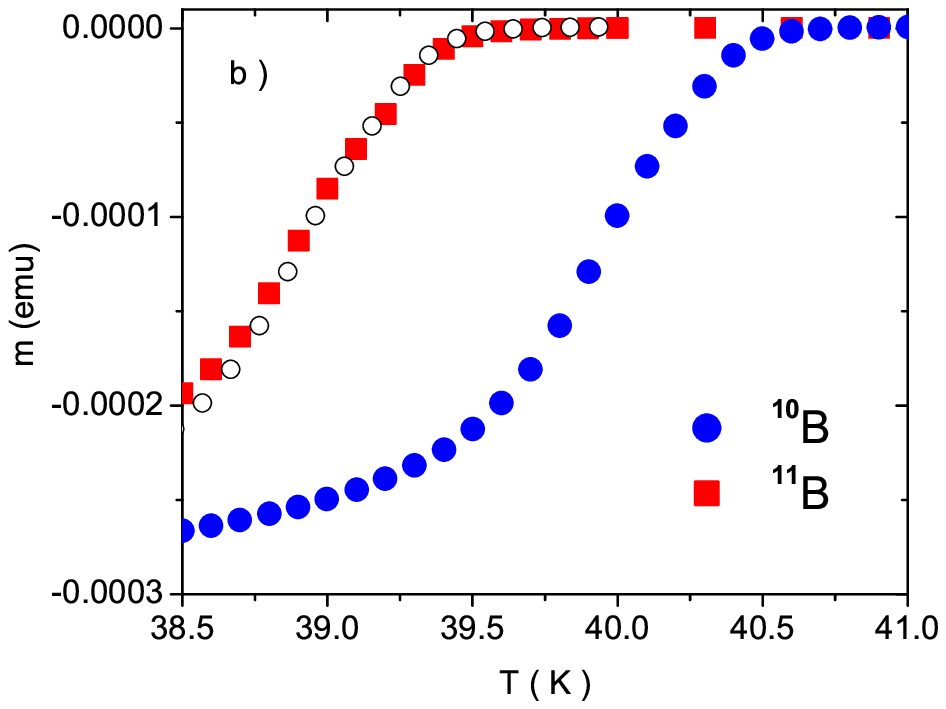}
\caption{Zero field cooled magnetization data at fixed magnetic
field \textit{vs}. $T$ near $T_{c}$ of a Mg$^{10}$B$_{2}$ and
Mg$^{11}$B$_{2}$ powder samples taken from Hinks \emph{et
al}.\protect\cite{hinks} (Fig.\ref{fig3}a) and Di Castro \emph{et
al}.\protect\cite{daniele} (Fig.\ref{fig3}b). The open circles are
the Mg$^{10}$B$_{2}$ data rescaled according to Eq.(\ref{eq4})
with $a=0.972$ (Fig.\ref{fig3}a) and $a=0.974$ (Fig.\ref{fig3}b).
} \label{fig3}
\end{figure}

Next we turn to the boron isotope effect. In Fig.\ref{fig3} we
show the normalized zero field cooled magnetization data
\textit{vs}. $T$ near $T_{c}$ of a Mg$^{10}$B$_{2}$ and
Mg$^{11}$B$_{2}$ powder samples taken from Hinks \emph{et al}.
\cite{hinks} (Fig.\ref{fig3}a) and Di Castro \emph{et al}.\cite
{daniele} (Fig.\ref{fig3}b). The dots are the Mg$^{10}$B$_{2}$
data rescaled according to Eq.(\ref{eq4}) with $a=0.972$
(Fig.\ref{fig3}a) and $a=0.974$ (Fig.\ref{fig3}b). Apparently,
$\Delta M/M\simeq 0$ within experimental error. Thus

\begin{equation}
-\frac{\Delta T_{c}}{T_{c}}=\frac{\Delta V\left( T_{c}\right)
}{\Delta V\left( T_{c}\right) }+\frac{\Delta f\left( \gamma \left(
T_{c}\right) \right) }{f\left( \gamma \left( T_{c}\right) \right)
}=1-a\simeq 0.028,0.026, \label{eq7}
\end{equation}
close to the value emerging from the pressure effect at $1.13$ GPa
(Eq.(\ref {eq6})) and the estimate of Hinks \emph{et
al}.\cite{hinks}. Since there is no conclusive evidence for any
significant lattice constant change for this isotope
exchange\cite{hinks2}, the volume change appears to be negligibly
small so that $\left| \Delta V\left( T_{c}\right) /V\left(
T_{c}\right) \right| <<\Delta f\left( \gamma \left( T_{c}\right)
\right) /f\left( \gamma \left( T_{c}\right) \right) $ holds.
Hence, in analogy to the pressure effect, the reduction of $T_{c}$
reflects essentially an increase of the anisotropy and with
$\Delta \gamma \left( T_{c}\right) /\gamma \left( T_{c}\right)
=\Delta \lambda _{c}\left( T_{c}\right) /\lambda _{c}\left(
T_{c}\right) -\Delta \lambda _{a}\left( T_{c}\right) /\lambda
_{a}\left( T_{c}\right) $ a change of the $c$-axis and / or
in-plane magnetic penetration depths. Within the microscopic
mean-field scenario\cite{dahm,gorkov,tsv} it implies a
renormalization of the Fermi surface topology due to thermal
fluctuations modified by isotope exchange or applied pressure.

To summarize, we have shown that in the two-band superconductor
MgB$_{2}$ the relative change of the transition temperature upon
Boron isotope exchange or applied pressure mirrors near $T_{c}$
essentially that of the anisotropy. Because this property stems
from 3D- Gaussian or 3D-XY thermal fluctuations and the
experimental fact that close to $T_{c}$ the magnetization scales
within experimental error as $^{i}M\left( T\right) =$ $^{j}M\left(
aT\right) $, where $^{i\neq j}M$ denotes the magnetization for
different isotopes or taken at different pressures, it appears to
be a universal property of anisotropic type II superconductors.
Indeed, in a variety of cuprate superconductors the relative
change of $T_{c}$ upon isotope exchange was also found to mirror
that of the anisotropy\cite{tsisom}. Thus, in anisotropic type II
superconductors the pressure or isotope exchange induced variation
of $T_{c}$ does not single out the mechanism mediating
superconductivity but mirrors predominantly the change of the
anisotropy, which puts a stringent constraint on microscopic
treatments.

\acknowledgments The authors are grateful to R. Khasanov and H.
Keller for fruitful discussions and comments on the subject
matter. This work was partially supported by the Swiss National
Science Foundation.

\end{document}